\documentclass{article}
\usepackage{spconf,amsmath,graphicx}
\usepackage{amssymb,amsfonts}
\usepackage{epsfig}
\usepackage{adjustbox}
\usepackage{cite}
\usepackage{todonotes}
\usepackage{hyperref}
\usepackage{booktabs}
\usepackage{graphicx}
\usepackage{float}
\usepackage{multirow}
\usepackage{arydshln}
\usepackage{enumitem}

\newcommand{\comment}[1]{}

\comment{
\let\oldbibliography\thebibliography
\renewcommand{\thebibliography}[1]{%
  \oldbibliography{#1}%
  \setlength{\itemsep}{0pt}%
}}

\usepackage[skip=5pt]{caption}

\hypersetup{
    colorlinks=true,
    linkcolor=olive,
    citecolor=magenta,      
    urlcolor=cyan,
    pdftitle={SincConv in SE}
    }

\title{What do neural networks listen to? Exploring the crucial bands in Speech Enhancement using Sinc-convolution}
\name{%
    Kuan-Hsun Ho$^{\dagger}$%
    \qquad Jeih-weih Hung$^{\star}$%
    \qquad Berlin Chen$^{\dagger}$%
}
\address{%
    $^{\dagger}$ \small{Department of Computer Science and Information Engineering, National Taiwan Normal University, Taipei, Taiwan} \\%
    $^{\star}$ \small{Department of Electrical Engineering, National Chi Nan University, Nantou, Taiwan}
}

\begin{document}
%
\maketitle
\begin{abstract}
This study introduces a reformed Sinc-convolution (Sincconv) framework tailored for the encoder component of deep networks for speech enhancement (SE). The reformed Sincconv, based on parametrized sinc functions as band-pass filters, offers notable advantages in terms of training efficiency, filter diversity, and interpretability. The reformed Sinc-conv is evaluated in conjunction with various SE models, showcasing its ability to boost SE performance. Furthermore, the reformed Sincconv provides valuable insights into the specific frequency components that are prioritized in an SE scenario. This opens up a new direction of SE research and improving our knowledge of their operating dynamics.
\end{abstract}
\begin{keywords}
Speech Ehancement, Sinc-convolution, Interpretability
\end{keywords}
\section{Introduction}
\label{intro}
Speech enhancement (SE) aims to ameliorate the quality and intelligibility of speech signals afflicted by noise interference, reverberations, and environmental disturbances. This pivotal field addresses the challenges posed by degraded speech in diverse contexts, including online meetings \cite{meeting}, hearing aids \cite{hearaid}, and speech recognition \cite{rasr}. 
Recent research attests DNN-fueled methods as superior alternatives to traditional statistical approaches, particularly excelling in combating non-stationary noise scenarios \cite{dnnse1, dnnse2, tstnn, demucs, metricgan+, fsn+, manner}. 

DNN-fueled SE methods are categorized broadly into spectral- and time-domain approaches based on the feature extraction from the raw waveform. 
In spectral-domain approaches, most employ hand-crafted features produced by deterministic filterbanks such as Short-time Fourier Transform (STFT) \cite{metricgan+, fsn+}, Gammatone filter (GTF) \cite{gtfse, gtfss}, or Mel Frequency Cepstral Coefficients (MFCC) \cite{mfcc}. 
Neural networks devised in this category either directly apply the phase of noisy input or aim for complex-valued masks, resulting in suboptimal results or challenging estimation \cite{dnnse2, tstnn}. On the other hand, time-domain approaches obtain speech representations through learning filterbank-like structures with uni-dimensional convolutional layers (Conv1d) to process magnitude and phase information simultaneously \cite{demucs, manner, convtasnet}. However, Conv1d requires a small kernel size (less than 16 sample points) to converge, translating to incongruous weights with limited interpretability. This implicit process makes sense for the neural network yet does not appeal to human intuition \cite{kernelize}.

The encoder-decoder pair is one of the most critical parts in the current mainstream of SE algorithms, especially the encoder, which not only deals with high-order information dwelling in input waveforms but profoundly affects whether the gradient vanishes as to the learned network. As such, we seek more freedom in learning filter weights as time-domain approaches while retaining interpretability as spectral-domain approaches. 
This objective compels us to investigate parametric filters. In particular, Sinc-convolution (Sinc-conv) \cite{sincnet} piques our interest.

Sinc-conv convolves the waveform with parametrized sinc functions that serve as band-pass filters, and the low and high cutoff frequencies are the sole parameters that necessitate learning during network optimization. Sinc-conv has been proven effective in speaker identification \cite{sincnet}, speech recognition \cite{kernelize, asr-sinc}, and mispronunciation detection \cite{capt-sinc}. Nonetheless, these tasks are cast as classification, mapping from higher-dimension (based on the framed signals) to lower-dimension representations (based on the classes). Therefore, we recast the formation of the original Sinc-conv, facilitating a more effective optimization suitable for SE. Additionally, the re-formation expands the diversity of the filter types by incorporating low-pass and high-pass options. Comprehensive experiments verify that we can boost the state-of-the-art by integrating our reformed Sinc-conv, while being capable of dissecting what neural networks pay attention to when performing SE.

\comment{\todo[inline]{Contribution}}

\section{Reforming Sinc-conv}
\label{modifed}

\comment{This section introduces the newly presented single-channel masking-based SE framework that employs Sinc-conv.} Generally speaking, a masking-based SE method briefly consists of the following steps: First, given a noisy utterance $x \in \mathbb{R}^T$, the encoder transforms $x$ to gain analytic speech feature $\tilde{x} \in \mathbb{R}^{N\times T'}$. \comment{Here, $T$ is the length of $x$, and $N$ and $T'$, respectively, denote the frame-wise feature size and the total number of frames of $\tilde{x}$.} Afterward, $\tilde{x}$ is used to learn a mask $m \in \mathbb{R}^{N\times T'}$ that suppresses the unwanted noise component by element-wise multiplying $m$ with $\tilde{x}$. The multiplied output then feeds the decoder to obtain an estimate of the clean utterance $\hat{x} \in \mathbb{R}^T$. The ultimate objective is to improve discourse quality and intelligibility while reducing noise. The method proposed in this study centers on the development of an encoder-decoder pair utilizing a reformed Sinc-conv.

\subsection{Formulation}
\label{formulation}

Sinc-conv entails convolving an input signal with $N$ filters, each having a finite impulse response (FIR) $h_i[n]$ that includes a truncated sinc function. The output of each filter can be expressed by: 
\begin{equation}\label{conv}
   \tilde{x}_i[n]=x[n] \ast h_i[n]=\sum^{L-1}_{l=0} x[n-l] h_i[l], 
\end{equation}
where an odd number $L=2M+1$ denotes the number of taps. For simplicity of discussion, we omit the filter index $i$ without losing generality. Each filter $h[n]$ is to approximate an ideal band-pass (magnitude) frequency response as:
\begin{equation}
    H(j\omega) = \text{rect}\left(\frac{\omega}{2\omega_{c_2}}\right)-\text{rect}\left(\frac{\omega}{2\omega_{c_1}}\right),
\end{equation}
where $\omega_{c_1}$ and $\omega_{c_2}$ represent the low and high cutoff frequencies, respectively, and $\text{rect}(x)$ is a rectangular function defined by $\text{rect}(x) = 1$ for $-0.5 \leq x \leq 0.5$ and $\text{rect}(x) = 0$ elsewhere. The inverse Fourier transform of $H(j\omega)$ is :
\begin{equation}
 \tilde{h}[n] = \frac{\omega_{c_2}}{\pi}\text{sinc}\left(\omega_{c_2}n\right) - \frac{\omega_{c_1}}{\pi}\text{sinc}\left(\omega_{c_1}n\right),
\end{equation}
where $-\infty < n < \infty$ and the sinc function is defined by $\text{sinc}(x)=\frac{\sin(x)}{x}$.  $\tilde{h}[n]$ is then delayed by $M$ and truncated, to be length of $L$, producing 
a causal FIR filter $\hat{h}[n] = \tilde{h}[n-M]$ for $0\leq n\leq L-1$ and $\hat{h}[n] =0$ elsewhere. Convolving with such a filter extracts the components of $x[n]$ that approximately fall within the bandwidth $[\omega_{c_1}, \omega_{c_2}]$.

However, in the original setting \cite{sincnet}, the parameters $\omega_{c_1}$ and $\omega_{c_2}$ are determined during DNN optimization, indicating that the corresponding range can run over $[0, \infty]$. This raises two concerns: 
First, updating weights is strenuous. In fact, many issues have reported that the learned filterbanks resemble their initial appearance\footnote[1]{\url{https://github.com/mravanelli/SincNet/issues?q=frequency+response}}; in other words, the learned frequencies have not been significantly altered.
Second, the Nyquist theorem might be contravened — the cutoff frequencies $\omega_{c_1}$ and $\omega_{c_2}$ should be within the range $[0, \omega_{N}]$, where $\omega_{N}$ denotes the Nyquist rate.
\comment{
\begin{enumerate}[leftmargin=15px]
\setlength\itemsep{0em}
\item Updating weights is strenuous. In fact, many issues have reported that the learned filterbanks resemble their initial appearance\footnote[1]{\url{https://github.com/mravanelli/SincNet/issues?q=frequency+response}}; in other words, the learned frequencies have not been significantly altered. 
\item The Nyquist theorem might be contravened — the cutoff frequencies $\omega_{c_1}$ and $\omega_{c_2}$ should be within the range $[0, \omega_{N}]$, where $\omega_{N}$ denotes the Nyquist rate. \comment{and is one half of the sample rate $\omega_s$.}
\end{enumerate}
}

Consequently, we recast the formation as:
\begin{equation}
    \begin{cases}
     \omega_{c_1}=\alpha_{c_1}\cdot \omega_{N} \\
     \omega_{c_2}=\alpha_{c_2}\cdot \omega_{N},
    \end{cases}
\end{equation}
and to ensure that $0<\omega_{c_1}<\omega_{c_2}<\omega_{n}$, the intermediate parameters (normalized frequencies relative to the Nyquist rate), $\alpha_{c_1}$ and $\alpha_{c_2}$, are calculated as:
\begin{equation}
    \begin{cases}
     \alpha_{c_1} = \min(\min(|\alpha_{c_1}^{raw}|,|\alpha_{c_2}^{raw}|), 1) \\
     \alpha_{c_2} = \min(\max(|\alpha_{c_1}^{raw}|,|\alpha_{c_2}^{raw}|), 1),
    \end{cases}
\end{equation}
where the raw normalized frequencies, $\alpha_{c_1}^{raw}$ and $\alpha_{c_2}^{raw}$, are the only parameters involved in training optimization. This way, updating $\alpha_{c_1}^{raw}$ and $\alpha_{c_2}^{raw}$ will be more influential in altering the cutoff frequencies compared to directly updating $\omega_{c_1}$ and $\omega_{c_2}$, thereby encouraging the pursuit of optimal solutions across multiple bands. Furthermore, we apply a learnable band gain (BG) $\beta_i \ge 0$ for each filter $h_i[n]$ and use a Hamming window $w[n]$ for mitigating the truncation effect, resulting in the final filter for the $i$-th band being expressed as:
\begin{equation}
    h_i[n]=\beta_i \cdot \hat{h}_i[n]\cdot w[n], 
\end{equation}
where $\beta_i$ is initialized to 1. 
In our implementation, an optional layer-wise normalization is first applied to 
the convolved output $\tilde{x}_i[n]$ in Eq. (\ref{conv}) then multiplying $\beta_i$. 

As for the decoder counterpart, we examine three options:
\begin{enumerate}[leftmargin=15px]
\setlength\itemsep{0em}
\item A learnable transposed Conv1d (T-Conv1d). 
\item The linear combination of the masked outputs. We merge $N$ channels into 1 through a linear combination with weights, $softmax(\{ \gamma_i \}_{i=1}^N)$, where each $\gamma_i$ is learnable. 
\comment{By doing so, we assume the importance of each band may not be identical to that of the noisy signal. }
\item The pseudo-inverse (pinv) of the learned filterbanks. 
\end{enumerate}

\subsection{Cutoff Frequencies Initialization}
\label{initialization}

Setting proper initial cutoff frequencies for filters $\hat{h}_i[n]$ in the encoder is crucial to boosting and improving optimization. Therefore, we adopt three initialization strategies for the raw normalized cutoff frequency pair $(\alpha_{c_1}^{raw}, \alpha_{c_2}^{raw})$:
\begin{enumerate}[leftmargin=15px]
\setlength\itemsep{0em}
\item	Sampling from a Uniform distribution. The frequency pair is yielded by drawing samples from a standard uniform distribution $\mathcal{U}[0,1]$. We set this as the default method.
\item	Sampling from a learned formant distribution: In \cite{sincnet}, it has been demonstrated that the cumulative frequency response (CFR) curve of the learned filters is analogous to a formant distribution. We interpolate and normalize this curve into a probability mass function (PMF) and then sample frequency pairs from that PMF.
\item	Sampling from the Mel-scale, as in \cite{sincnet}. \comment{The only distinction is that we use normalized frequencies relative to the Nyquist frequency.}
\end{enumerate}

\subsection{Attribute}
\label{attribution}

As stated in \cite{sincnet}, Sinc-conv has multiple remarkable attributes, including faster convergence and fewer parameters than a conventional Conv1d. Based on the explicit expression of a sinc function, Sinc-conv enables the encoder network to focus on tuning fewer parameters with physical meaning (the cutoff frequencies), accelerating convergence. However, when considering the presented re-formation and task orientation, there would be more potential attributes and benefits.

Considering the magnitude of the update and the learning rate sensitivity, minor updates are generally easier to manage during training since they lead to a smoother convergence toward the optimal solution.
Moreover, since cutoff frequencies are permitted to be more flexible, it creates a greater diversity of filter types. The learned filters are either low-, high-, or band-pass. Furthermore, the BGs reveal distinct levels of influence for each band in SE. \comment{The reasoning behind this is that different frequency components are unequally significant to human perception \cite{mulca1, mulca2}, and that the energy of a speech utterance is typically distributed in a non-uniform fashion across frequencies.}
Last but not least, due to its symmetric coefficients, each filter $\hat{h}_i[n]$ has a linear phase response that indicates no phase distortion due to time delay, which is crucial in generative task like SE. \comment{This attribute is crucial in SE, as the signal shape is paramount for maintaining high fidelity in the filtered audio.}

\section{Experiment}
\label{experiment}

\subsection{Setup}
\label{setup}

\textbf{Dataset.} 
To assess the effectiveness of our proposed method, we conducted experiments on the VoiceBank-DEMAND \cite{vcbk-dm} benchmark dataset. This widely used open-source corpus for SE includes 11,572 pre-synthesized training utterances (from 28 speakers) with ten types of noise from the DEMAND database \cite{dm} at four SNR values (0, 5, 10, and 15 dB). The test set comprises 824 utterances (from 2 speakers) contaminated by five types of noise at SNR values of 2.5, 7.5, 12.5, and 17.5 dB. We reserved approximately 200 training set utterances for validation. All speech data has a 16 kHz sample rate.

\noindent\textbf{Model.} 
For our reformed Sinc-conv framework, we set $N=80$ pairs of cutoff frequencies and the kernel size $L$ as 251. \comment{To re-implement the experiments, note that $g[n]$ and $w[n]$ have different periodicity, i.e., $g[n]=g(n/f_s)$ and $w[n]=w(n/L)$.}
We validate the framework as the encoder-decoder component of two renowned SE models: Conv-TasNet \cite{convtasnet} and MANNER \cite{manner}. In Conv-TasNet, we set the baseline configuration with default hyperparameters ($N, L, B, H, P, X, R$) as specified in \cite{convtasnet}.
The Sinc-conv method is evaluated using the Conv-TasNet model with revised hyperparameters $B=120$ and $H=256$ while keeping the rest of the hyperparameters unchanged. Such revised Conv-TasNet reduces the total parameter count by 46\%. The loss function uses the negative scale-invariant source-to-noise ratio (SI-SNR).
On the other hand, MANNER, which is considered state-of-the-art, demonstrates more effectiveness in SE. We incorporate the presented Sinc-conv into the MANNER framework to assess its potential to boost SE behavior. However, our evaluation is restricted to a reduced-scale version of MANNER, as mentioned in \cite{manner}, due to limited computational resources. The initial hidden size in the MANNER module is designated as $N=80$, and all other hyperparameters and the loss function remain consistent with \cite{manner}. \comment{Afterwards, the hidden size for the subsequent level will be doubled to 160, and this pattern will continue.}

\noindent\textbf{Evaluation metrics.}
Perceptual Evaluation of Speech Quality (PESQ) and Short-time objective intelligibility (STOI) are often used as objective metrics for evaluating speech quality and intelligibility. We use both PESQ and STOI to compare the performance. \comment{Furthermore, we prepare an ASR system to recognize the enhanced speech and calculate the word error rate (WER). The ASR model is a Kaldi-based hybrid DNN-HMM acoustic model trained with lattice-free MMI objective function using all the clean utterances in VoiceBank-DEMAND.} Furthermore, SI-SNR is also reported to quantify noise reduction, and higher SI-SNR values indicate better noise suppression.

\subsection{Result}
\label{result}

\begin{table}[tb]
\caption{The comparison in Conv-TasNet. "ln", "sc", and "lc" stands for "learnable", "Sinc-conv", and "linear combination". The "org. form" implies using the original Sinc-conv.}
\label{tab:convtasnet}
\centering
\resizebox{6.8 cm}{!}{%
\begin{tabular}{cccccccc}
\hline
 & \multicolumn{2}{c}{Encoder} & \multicolumn{2}{c}{Decoder} & \multirow{2}{*}{PESQ} & \multirow{2}{*}{STOI(\%)} & \multirow{2}{*}{SI-SNR} \\
                  & ln         & sc         & ln         & lc         &  &  &  \\ \hline
\textit{Baseline}                  & \checkmark &            & \checkmark &            & 2.57 & 93.1 & 18.32 \\ \hline
\textit{Proposed}                  &            & \checkmark & \checkmark &            & \textbf{2.65} & \textbf{93.6} & 18.37 \\
\textit{}                          &            & \checkmark &            & \checkmark & 2.61 & 93.5 & \textbf{18.63} \\ \hdashline
\textit{\small{(org. form)}}         &            & \checkmark & \checkmark &            & 2.56 & 93.1 & 18.01 \\ \hline
\end{tabular}%
}
\end{table}

Table ~\ref{tab:convtasnet} shows the results of testing the ConvTasNet module with various modifications. Our reformed Sinc-conv performs better, even when employing a simple linear combination as the decoder. Contrarily, using original Sinc-conv yields outcomes merely comparable to the baseline Conv-TasNet. However, note that using Sinc-conv can substantially reduce 46\% in model parameters, which is quite impressive. This demonstrates the capability of Sinc-conv to reduce the dimension of features, in this case, from 512 to 80.

\begin{table}[tb]
\caption{The comparison in MANNER. "fm init." and "mel init." stands for "formant" and "Mel-scale initialization".}
\label{tab:manner}
\centering
\resizebox{8.1 cm}{!}{%
\begin{tabular}{ccccccccc}
\hline
\multirow{2}{*}{} & \multicolumn{2}{c}{Encoder} & \multicolumn{3}{c}{Decoder} & \multirow{2}{*}{PESQ} & \multirow{2}{*}{STOI(\%)} & \multirow{2}{*}{SI-SNR} \\
                           & ln         & sc         & ln         & lc         & pinv       &    &    &    \\ \hline
\textit{Baseline}            & \checkmark &            & \checkmark &            &            & 2.91 & 94.5 & 18.70 \\ \hline
\textit{Proposed}            &            & \checkmark & \checkmark &            &            & \textbf{3.03} & \textbf{95.0} & \textbf{18.91} \\
\textit{}                    &            & \checkmark &            & \checkmark &            & 2.96 & 94.6 & 18.81 \\
\textit{}                    &            & \checkmark &            &            & \checkmark & 2.81 & 94.2 & 18.19 \\ \hdashline
\textit{\small{(fm init.)}}  &            & \checkmark & \checkmark &            &            & 2.96 & 94.6 & 18.61 \\
\textit{}                    &            & \checkmark &            & \checkmark &            & 2.89 & 94.4 & 18.62 \\ \hdashline
\textit{\small{(mel init.)}} &            & \checkmark & \checkmark &            &            & 2.85 & 94.3 & 18.53 \\
\textit{}                    &            & \checkmark &            & \checkmark &            & 2.80 & 94.1 & 18.35 \\ \hdashline
\textit{\small{(org. form)}} &            & \checkmark & \checkmark &            &            & 2.75 & 94.1 & 18.20 \\ \hline
\end{tabular}%
}
\end{table}

Table ~\ref{tab:manner} reports the results experimented on MANNER in various forms. Our reformed Sinc-conv, when paired with either T-Conv1d or linear combination as the decoder, performs better than the baseline. However, the pinv decoder fails probably due to the ambiguity in its interpretation as a pseudo matrix inverse of learned filterbanks. A similar finding is also reported in \cite{convtasnet, gtfss}.
As for the initialization method, uniform sampling performs the best, followed by formant distribution sampling and then Mel-scale frequencies. Uniform sampling introduces stochasticity, which stimulates the exploration of different frequencies, potentially helping Sinc-conv try out more band locations. Comparatively, formant sampling leverages the prior knowledge obtained from speaker identification tasks, yet the resulting sampled cutoff frequencies are not necessarily feasible for SE purposes. 
The inferior results when employing Mel-scale frequency initialization for both the original and reformed Sinc-conv demonstrate the limitations of this initialization method. Despite the better performance in the reformed case, this setback emphasizes the insufficiency of  relying solely on band-pass filters.

An ablation study is conducted and shown in Tab.~\ref{tab:ablation}. Surprisingly, increasing the kernel size (filter length) $L$ does not gain a more favorable outcome, even though it leads to higher frequency resolution without growing the amounts of parameters. As noted in \cite{dprnn}, we believe there are certain underlying principles in assigning the kernel size to each layer, but we leave this to future efforts. On the contrary, increasing the number of cutoff frequency pairs $N$ is beneficial. 
Finally, we examine the impact of BGs and layer normalization. While the BGs contribute less, layer normalization is highly advantageous for the provided framework for its significance in achieving stable and faster convergence.

\begin{table}[tb]
\caption{The ablation study conducted on "reformed Sinc-conv + MANNER + T-Conv1d."}
\label{tab:ablation}
\centering
\resizebox{6.3 cm}{!}{%
\begin{tabular}{ccccccc}
\hline
\multicolumn{4}{c}{Components} & \multirow{2}{*}{PESQ} & \multirow{2}{*}{STOI(\%)} & \multirow{2}{*}{SI-SNR} \\
$L$    & $N$     & $\beta_i$   & norm &  &  &  \\ \hline
251    & 80      & \checkmark & \checkmark & 3.03 & 95.0 & 18.91 \\
1025   & 80      & \checkmark & \checkmark & 2.73 & 94.0 & 18.02 \\
251    & 120     & \checkmark & \checkmark & 3.10 & 95.2 & 19.17 \\
251    & 80      & \checkmark &            & 2.74 & 94.1 & 18.16 \\
251    & 80      &            & \checkmark & 2.95 & 94.7 & 18.74 \\ \hline
\end{tabular}%
}
\end{table}

\section{Discussion}
\label{discussion}

\comment{
\begin{figure}[tb]
\begin{minipage}[b]{1.0\linewidth}
  \centering
  \centerline{\epsfig{figure=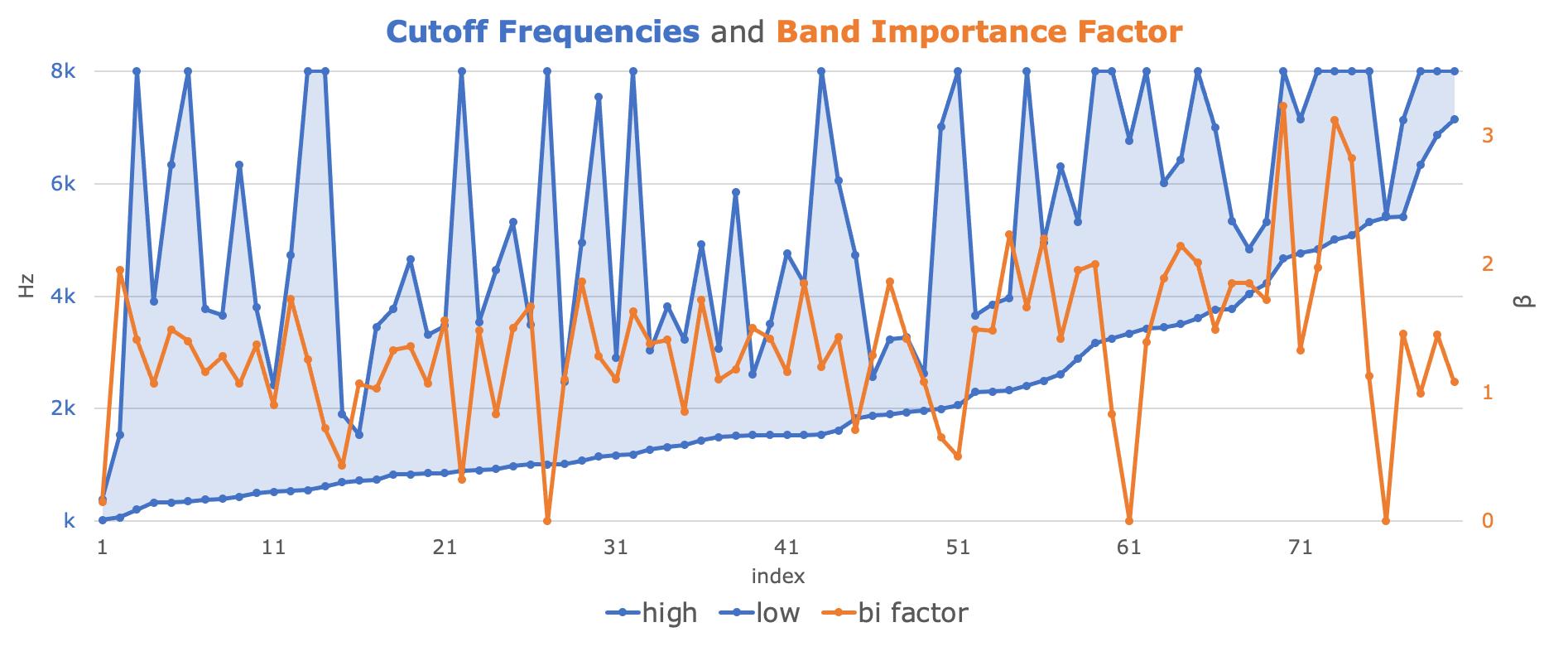, width=8.5cm}}
   \centerline{\small{(a) "ln" as decoder.}}\medskip
\end{minipage}
\begin{minipage}[b]{1.0\linewidth}
  \centering
  \centerline{\epsfig{figure=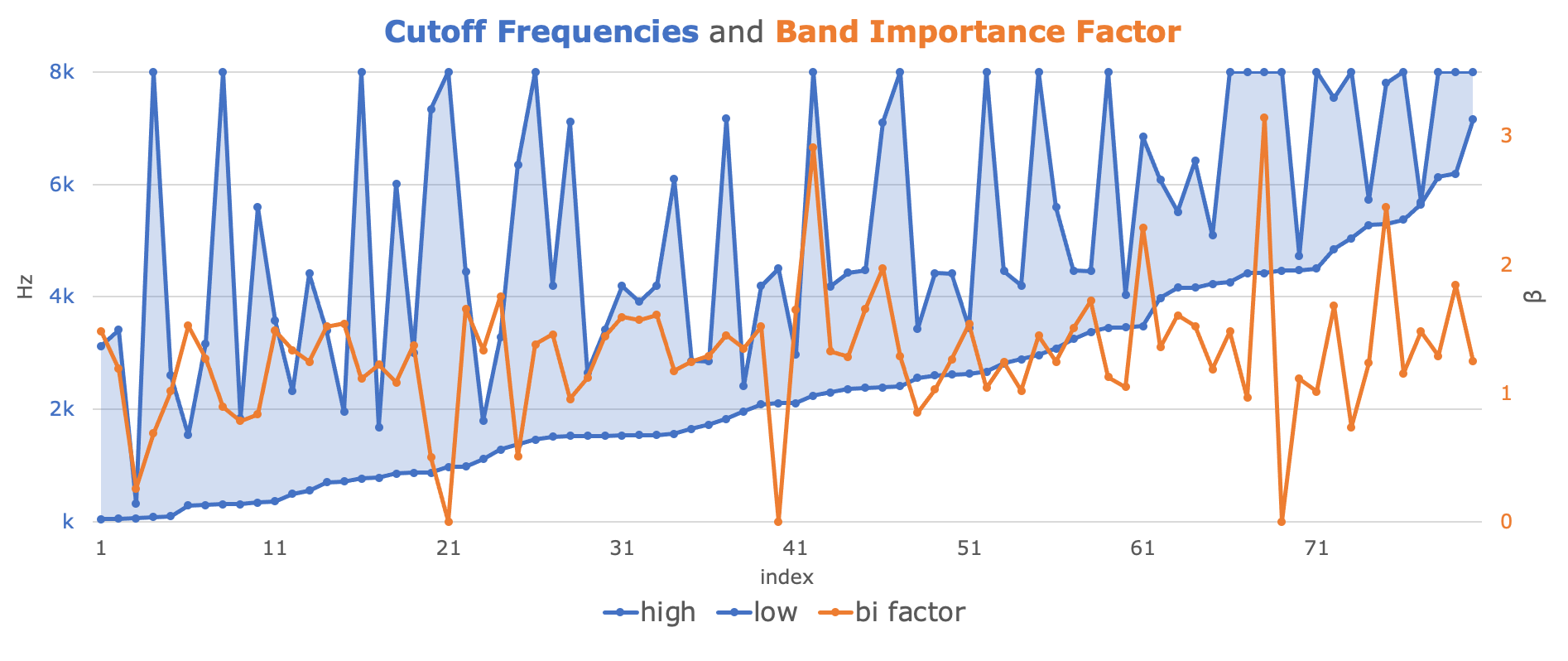, width=8.5cm}}
  \centerline{\small{(b) "lc" as decoder.}}\medskip
\end{minipage}
\caption{The cutoff frequencies and BGs, sorted by the lower frequencies.}
\label{fig:cutoffs}
\end{figure} 
}

\begin{figure}[tb]
    \centerline{\includegraphics[width=8.1 cm]{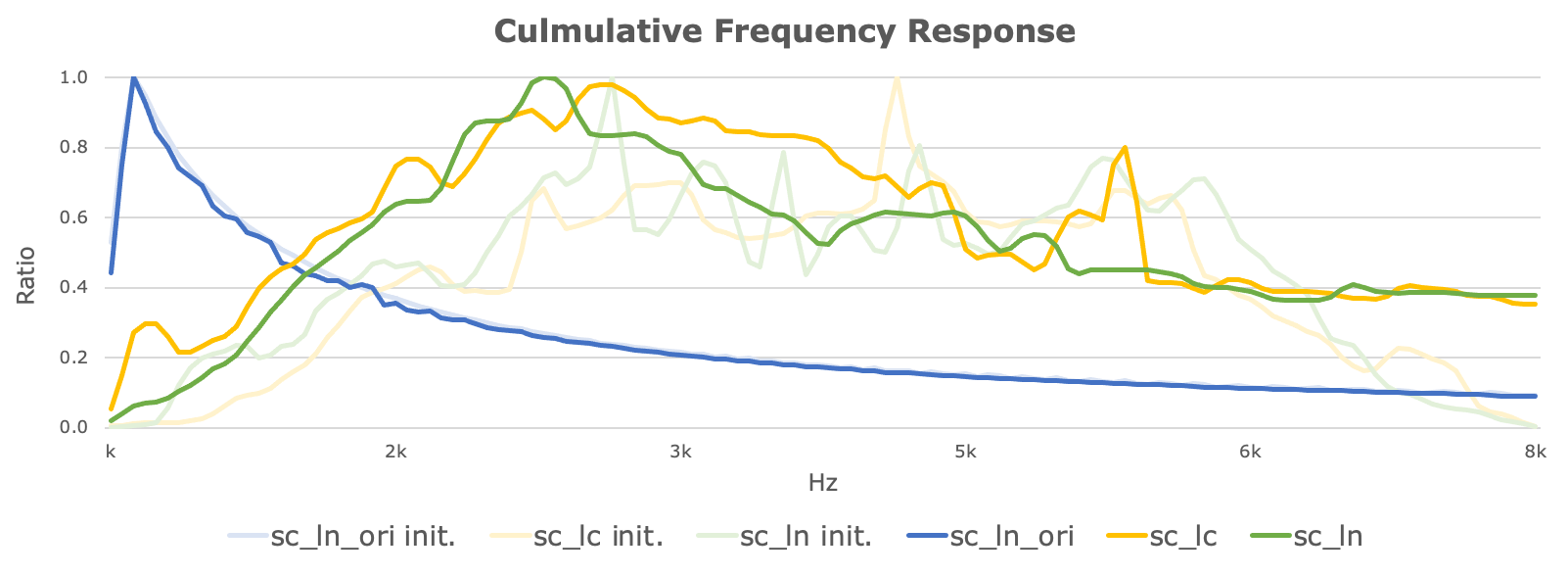}}
    \caption{CFR.}
    \label{fig:cumfr}
\end{figure}

\comment{This section dissects Sinc-conv to visualize how the magic is done.} The CFR curves derived from different encoder-decoder pairs, namely: 1) "reformed Sinc-conv + T-Conv1d", 2) "reformed Sinc-conv + linear combination," and 3) "original Sinc-conv + T-Conv1d", along with their corresponding initialization values, are depicted in Figure ~\ref{fig:cumfr}. \comment{Note that the spikes in the curves are caused by two rectangular filters aliasing, which is quite normal (?).} 

First, we can see that the original Sinc-conv does not significantly alter the CFR relative to the initial one. Contrarily, the reformed Sinc-conv exhibits a distinct variation in CFR as a result of its higher flexibility. Second, the reformed Sinc-conv places greater emphasis on medium frequencies, while the original Sinc-conv prioritizes low frequencies. Apparently, the CFR difference is the primary factor for the disparity in performance. While the Mel scale was initially designed for human auditory perception, it exhibits limitations when encountering noises. However, we must clarify that a lower CFR region does not necessarily imply that the corresponding frequencies are unimportant. In fact, the CFR curve serves as an indicator of the SE model's comprehensive listening and analytical efforts. 

To better understand the filter characteristics obtained by reformed Sinc-conv, we plot the cutoff frequencies and BGs of each filter in Fig~\ref{fig:cutoffs}. First, we find that the majority of the filters are band-pass, with some being high-pass, and a few being low-pass. Second, we notice that there are few BGs equal to zero, which operate similarly to a dropout layer and serve as a form of network regularization. Third, the lower frequency curves display exponential growth. We also observe the same growth when sorted by higher frequencies. This resembles the Mel Scale and mimics the way humans perceive sound. \comment{Due to the limitation in paper length, we leave various interesting analyses behind, notably the output mask and the linear combination decoder, as they operate on true signals composed of frequencies in a specific bandwidth. Nonetheless, significant progress in interpreting the focus of neural networks, a.k.a, black boxes, has been made through our study.}

\begin{figure}[tb]
    \centerline{\includegraphics[width=8.1 cm]{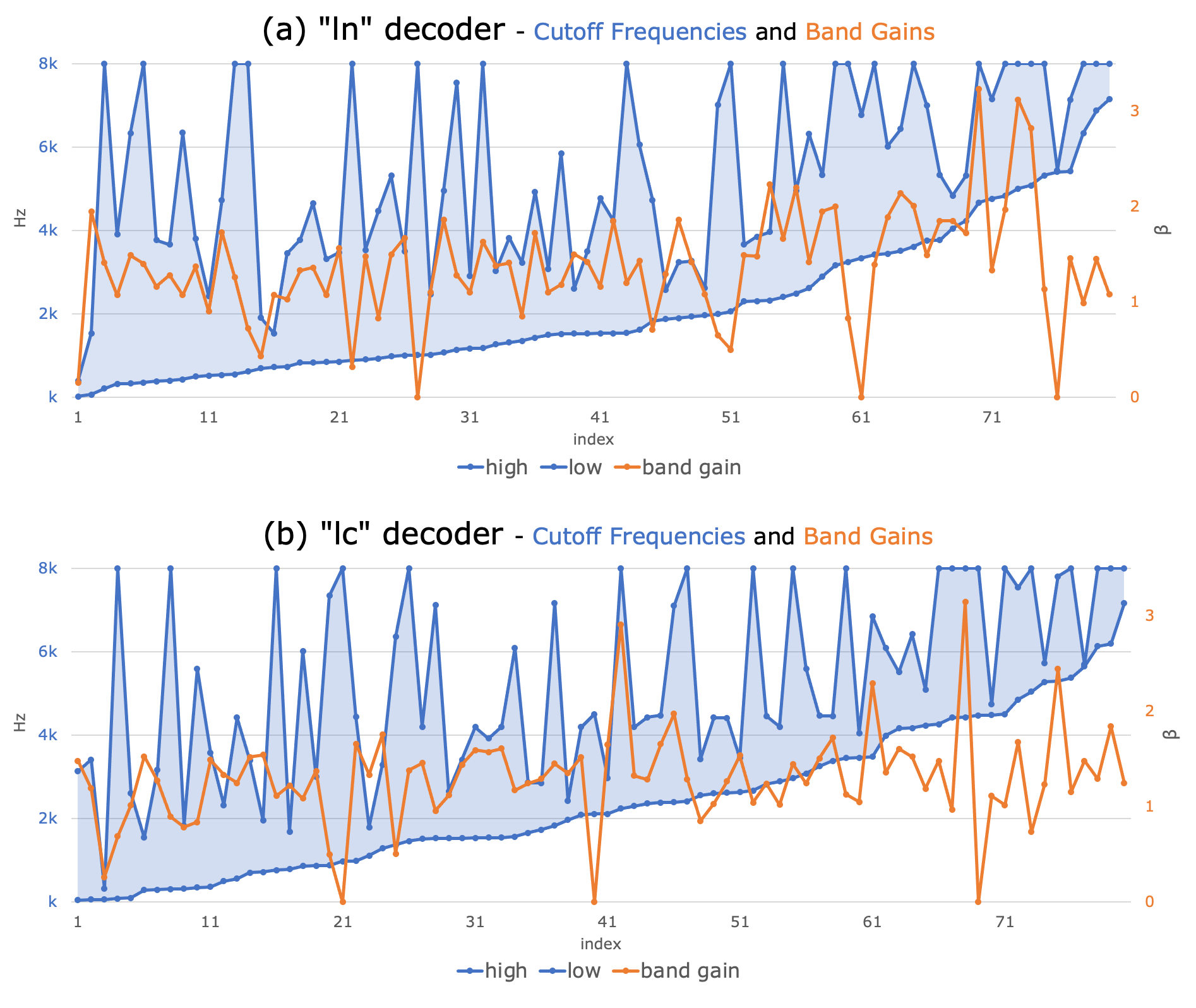}}
    \caption{The cutoff frequencies and BGs, sorted by the lower frequency.}
    \label{fig:cutoffs}
\end{figure}

\section{Conclusion}
\label{conclusion}

In this study, we propose a reformed Sinc-conv method that facilitates training efficiency and diversifies the filter types for the encoder of a deep SE network. The efficacy of the reformed Sinc-conv is evaluated with multiple SE models and configurations, clearly showcasing its potential to boost SE performance. By leveraging Sinc-conv, we can interpret what an SE network pursues to listen to when enhancing a noisy speech. To sum up, we have successfully fulfilled our goal of developing automatic learning filterbanks that exhibit high interpretability and effectiveness.

\section{Acknowledgement}

This work was supported in part by Realtek Semiconductor Corporation under Grant Number 112KK010. Any findings and implications in the paper do not necessarily reflect those of the sponsors.


\vspace{12pt}

\end{document}